\documentclass[conference, letterpaper]{IEEEtran}
\usepackage{cite}
\usepackage{amsmath,amssymb,amsfonts}
\usepackage{graphicx}
\usepackage{textcomp}
\usepackage{xcolor}
\usepackage{mathtools}
\usepackage{hhline}
\usepackage{array}
\newcolumntype{C}[1]{>{\centering\arraybackslash}m{#1}}
\usepackage[linesnumbered, ruled, noend]{algorithm2e}
\SetKwComment{Comment}{/* }{ */}

\SetCommentSty{mycommfont}
\usepackage[colorlinks=true]{hyperref}
\usepackage[english]{babel}
\addto\extrasenglish{

}
\begin{document}

\title{Depth-Optimal Addressing of \\ 2D Qubit Array with 1D Controls \\ Based on Exact Binary Matrix Factorization}
\author{\IEEEauthorblockN{Daniel Bochen Tan}
\IEEEauthorblockA{
\textit{University of California, Los Angeles}\\
bochentan@ucla.edu}
\and
\IEEEauthorblockN{Shuohao Ping}
\IEEEauthorblockA{
\textit{University of California, Los Angeles}\\
sp1831@ucla.edu}
\and
\IEEEauthorblockN{Jason Cong}
\IEEEauthorblockA{
\textit{University of California, Los Angeles}\\
cong@cs.ucla.edu}
}

\maketitle

\begin{abstract}
Reducing control complexity is essential for achieving large-scale quantum computing.
However, reducing control knobs may compromise the ability to independently address each qubit.
Recent progress in neutral atom-based platforms suggests that rectangular (row-column) addressing may strike a balance between control granularity and flexibility for 2D qubit arrays.
This scheme allows addressing qubits on the intersections of a set of rows and columns each time.
While quadratically reducing controls, it may necessitate more depth.
We formulate the depth-optimal rectangular addressing problem as exact binary matrix factorization, an NP-hard problem also appearing in communication complexity and combinatorial optimization.
We introduce a satisfiability modulo theories-based solver for this problem, and a heuristic, row packing, performing close to the optimal solver on various benchmarks.
Furthermore, we discuss rectangular addressing in the context of fault-tolerant quantum computing, leveraging a natural two-level structure.

\end{abstract}


\section{Introduction}

\begin{figure}[t]
\centerline{\includegraphics[scale=1]{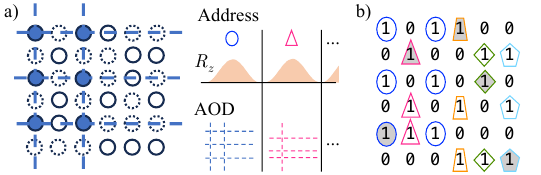}}
\caption{Rectangular addressing in neutral atom arrays.
a)~The experimental setup in Bluvstein et al.~\cite{bluvstein_logical_2023}: a 2D acousto-optic deflector (AOD, blue dashes) modulates another laser to realize $R_z$ gates on qubits at the AOD crossing points (colored dots).
Different qubits (uncolored dots) are addressed by changing the AOD signal.
Qubits not in the pattern (dash circles) should not be addressed.
b)~Rectangular partition of a).
Different markers distinguish 5 rectangles to partition the matrix.
The 5 filled markers indicate a fooling set.
}
\label{fig:atom}
\end{figure}

Our motivation stems from recent successful large-scale experiments on the neutral atom arrays platform \cite{bluvstein_logical_2023} that highlights the effect of reducing control complexity.
As depicted in \autoref{fig:atom}a, the acousto-optic deflector (AOD) illuminates a product of rows and columns.
Quantum gates, induced by specific pulses modulated by the AOD, address qubits at the row and column intersections.
AODs prove effective for implementing gates~\cite{nature22-graham-atom-array, bluvstein_logical_2023} and qubit movements~\cite{ebadi_quantum_2021, bluvstein_quantum_2022,  tan_qubit_2022, Tan2024compilingquantum}.

For a 2D array \(X \times Y\), a \textit{(combinatorial) rectangle} is a set of the form \(X' \times Y'\), where \(X' \subseteq X\) and \(Y' \subseteq Y\).
Specifying a rectangle requires \(|X| + |Y|\) bits (one bit for each row and each column), a significant reduction compared to \(|X| \cdot |Y|\) bits for all elements.
This quadratic reduction is maintained while preserving individual addressability, as a single element can still be considered a rectangle.

The coarser granularity of rectangular addressing may reduce control complexity at the cost of increasing depth.
A generic example problem is given by the matrix in \autoref{fig:atom}b, where the qubits to address are represented by the 1's.
This matrix can be partitioned into five rectangles, each designated by distinct markers.
Consecutively, each rectangle receives a modulated $R_z$ pulse through specific AOD configurations.
Minimizing the number of rectangles to partition arbitrary binary matrices becomes crucial.

In this paper, we consider the problem of achieving depth-optimal rectangular addressing, which we formulate as \textit{exact binary matrix factorization}. 
We present a satisfiability modulo theories (SMT, extension of SAT) formulation for this problem and an effective heuristic dubbed \textit{row packing}.
The combined algorithm, SAP (SMT and packing), finds high-quality heuristic solutions quickly and then iteratively approaches to the optimal solution.

To assess our methods, we generate three benchmark sets.
The first set comprises random matrices, the second includes matrices with known optimal solutions, despite the problem generally being NP-hard.
The third set is constructed to accentuate the gap between the partition number and the conventional matrix rank.

We also observe that for future fault-tolerant quantum computing (FTQC), the problem may exhibit a product structure.
This implies that we can solve limited-size problems on multiple levels and then combine the solutions.

The paper is structured as follows.
In \autoref{sec:background}, we provide a review of mathematical concepts related to this problem across various contexts and applications.
In \autoref{sec:algorithm}, we introduce our algorithms.
The construction of benchmarks and the evaluation results are presented in \autoref{sec:evaluation}.
We discuss the problem in the context of FTQC in \autoref{sec:ftqc}.
Finally, the conclusions and future directions are provided in \autoref{sec:conclusion}.

\section{Background \label{sec:background}}
The term we have adopted, \textit{rectangle}, is standard in communication complexity theory~\cite{yao_complexity_1979}, where the matrix in \autoref{fig:atom}b represents a binary function $g$ of two variables.
Alice has some $i$, Bob has some $j$, and our interest is determining the number of bits the two need to communicate to compute $g(i,j)$.
If $g=1$ uniformly on a rectangle, it is a \textit{1-monochromatic rectangle}.
The number of 1-monochromatic and 0-monochromatic rectangles to partition the whole matrix serves as a crucial lower bound for the communication complexity.
For an introduction to this topic, readers are referred to Kushilevitz \& Nisan~\cite{kushilevitz_communication_1997}.
Two results they cover are worth mentioning for later discussions.
First, there exists an alternative definition of a rectangle:
\begin{equation} \label{eq:def}
(i,j)\in R \text{ and } (i',j')\in R \Rightarrow (i,j')\in R.
\end{equation}
In this paper, we only focus on 1-monochromatic rectangles and we will refer to these as `rectangles' from now on.
The second important fact is that the partition number is lower bounded by the size of \textit{fooling sets}.
In our case, a fooling set $S$ consists of $(i,j)$ such that $g(i,j)=1$, and for any distinct pair $(i,j)$ and $(i',j')$ in $S$, $g(i',j) = 0$ or $g(i,j') = 0$.
Indeed, the shaded markers in \autoref{fig:atom}b identify such a fooling set of size 5, implying that our partition into 5 rectangles is optimal.
Fooling sets do not always guarantee a tight bound, e.g.,
\begin{equation} \label{eq:fooling}
\begin{array}{ll}
\text{3 rectangles are needed to partition} \\
\text{but the size of any fooling set is}\le 2
\end{array}
\ \ 
\begin{bmatrix}
1 & 1 & 0 \\
0 & 1 & 1 \\
1 & 1 & 1
\end{bmatrix}
\end{equation}

\begin{figure}[t]
\centerline{\includegraphics[scale=1]{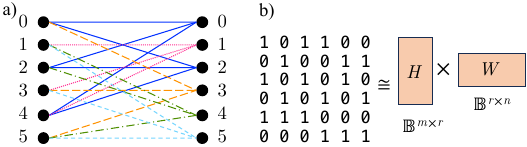}}
\caption{a)~Interpreting the matrix as the adjacency matrix of a bipartite graph, the rectangular partition problem becomes \textit{biclique partition} where the edges are partitioned to form complete bipartite subgraphs (different line types).
b)~Binary matrix factorization finds low-rank approximations $HW$ of the original matrix where $H$ and $W$ are also required to be binary.}
\label{fig:others}
\end{figure}

The problem has a graph-theoretic interpretation when considering the matrix as the adjacency matrix of a bipartite graph, as illustrated in \autoref{fig:others}a.
The left vertices correspond to the rows, while the right vertices correspond to the columns.
An edge exists between vertex $i$ on the left and vertex $j$ on the right if and only if element $(i,j)$ is 1 in the matrix.
Viewed in this way, a rectangle, seen as a set of edges, forms a \textit{biclique}, i.e., a complete bipartite graph.
For instance, the addressed sites in \autoref{fig:atom}a correspond to a complete \mbox{(3,2)-bipartite} subgraph in \autoref{fig:others}a, as denoted by the solid edges.
Therefore, the rectangular partition is equivalent to a biclique partition of a bipartite graph.
Reinterpreting the left vertices as sets and right vertices as objects, the biclique partition is finding a \textit{normal set basis} to decompose each set.
In our example, the basis is $\{\{0,2\}, \{1\}, \{3\}, \{4\}, \{5\}\}$, with the first set on the left decomposed into $\{0,2\}\sqcup \{3\}$.
The decision problem is proven NP-complete~\cite{jiang_minimal_1993}.
Even approximating the problem is NP-hard \cite{bein_clustering_2008, chalermsook_nearly_2014, chandran_parameterized_2017}.
Amilhastre et al.~\cite{amilhastre_complexity_1998} have characterized certain graph families where the problem can be efficiently solved.

The third perspective regarding a rectangular partition is through matrix factorization, as each rectangle precisely corresponds to a rank-1 submatrix.
In \autoref{fig:others}b, within a \textit{binary matrix factorization} (BMF), given a binary matrix $M\in \mathbb{B}^{m\times n}$ and an integer $r$, the objective is to minimize $\Vert M-HW \Vert$ where $H\in \mathbb{B}^{m\times r}$ and $W\in \mathbb{B}^{r\times n}$.
Note that $HW=\sum_{i=1}^{r}P_i$ where $P_i$ is the product of column $i$ in $H$ and row $i$ in $W$.
Each $P_i\in \mathbb{B}^{m\times n}$ is 1 on a combinatorial rectangle and 0 elsewhere, so it has rank 1.
The minimum $r$ for which $M-HW=0$ is the \textit{binary rank}, $r_\mathbb{B}$ of $M$.
In this case, $\sum_{i=1}^{r}P_i$ is an \textit{exact} binary matrix factorization (EBMF) of $M$.
In contrast to SVD, which provides the rank in $\mathbb{R}$, EBMF additionally requires $H$ and $W$ to be binary.
However, it is crucial to note that the additions in the matrix multiplication in EBMF is in $\mathbb{R}$, not in $\mathbb{B}$, e.g.,
\[
\begin{bmatrix}
0 & 1 & 1 \\
1 & 0 & 1 \\
1 & 1 & 0
\end{bmatrix}\ \overset{\text{EBMF}}{\neq}\ 
\begin{bmatrix}
1 \\
0 \\
1 
\end{bmatrix}
\begin{bmatrix}
1 & 1 & 0
\end{bmatrix}+
\begin{bmatrix}
1 \\
1 \\
0 
\end{bmatrix}
\begin{bmatrix}
1 & 0 & 1
\end{bmatrix}.
\]
If the addition were in \(\mathbb{B}\), the equality holds.
But in \(\mathbb{R}\), the top-left element appears in \textit{both} rectangles on the r.h.s., violating the disjointedness requirement of rectangular partitioning.
For binary matrices, we have a straightforward lower bound~\cite{watson_nonnegative_2016}:
\begin{equation} \label{eq:lowerbound}
\text{rank}_\mathbb{R}(M) \le r_\mathbb{B}(M)\ \ \forall M\in\mathbb{B}^{m \times n}.
\end{equation}
Zhang et al.~\cite{zhang_binary_2007} develop a BMF optimizer which is integrated into a well-known package NIMFA~\cite{Zitnik2012}.
However, since it is not designed for EBMF but to provide approximations given a fixed $r$, it does not perform well for our specific purposes.

\section{Algorithm \label{sec:algorithm}}
Given a matrix $M\in \mathbb{B}^{m\times n}$, we are interested in its exact binary matrix factorization (EBMF) $M=\sum_{i=0}^{r_\mathbb{B}-1} P_i$ where each $P_i\in \mathbb{B}^{m\times n}$ is 1 on a rectangle and 0 elsewhere.

Our SMT formulation encodes the problem: given $M$ and a number $b$, determine if $r_\mathbb{B}(M)\le b$.
When $r_\mathbb{B}$ is unknown, we query an SMT solver with decreasing values of $b$ to compute it.
Given the problem's complexity, the worst-case runtime is exponential to the size of $M$.
Hence, the key lies in establishing relatively tight bounds for $b$ to minimize SMT invocations.

Our approach, SAP (SMT and packing), is presented in \autoref{alg}.
First, our heuristic, \textit{row packing}, provides a valid EBMF, $P$.
Since $|P|$ is an upper bound of $r_\mathbb{B}(M)$, the SMT solving initiates with $b = |P| - 1$ and terminates when the SMT formula is unsatisfiable or when $b$ falls below $\text{rank}_\mathbb{R}(M)$, a lower bound as per \autoref{eq:lowerbound}.
$P$ is updated each time the SMT formula is satisfiable so that it retains to the best solution found thus far even if the process is prematurely interrupted.

\begin{algorithm}[t]
\caption{SAP (SMT and packing) EBMF}\label{alg}
\KwData{$M\in \mathbb{B}^{m\times n}$}
\KwResult{$P$, an EBMF of $M$ consisting of rectangles}
$P\gets$ row\_packing\_EBMF($M$)\Comment*[r]{\autoref{alg:packing}}
$b\gets |P| - 1$\;
\textit{formula} $\gets$ construct\_SMT\_formula($M$, $b$)\;
\While{$b\ge \text{real rank of } M $}{
  \eIf{formula is satisfiable}{
    $P \gets$ readout\_solution( \textit{formula})\;
    $b\gets b-1$\;
    \textit{formula} $\gets$ narrow\_down\_depth( \textit{formula}, $b$)\;
  }{
  \textbf{break}\;
  }
}
\end{algorithm}

\subsection{SMT Formulation \label{ssec:SAT}}
Fundamentally, we want to compute a function $f: E \to P$, where $E$ comprises the 1's in the matrix, and $P$ contains the rectangles.
This definition offers the convenience of inherently ensuring the disjointedness of the rectangles.
Furthermore, the constraints needed to enforce the validity of $f$ in specifying rectangles can be expressed using first-order logic and equality between function values.
This closely aligns with the uninterpreted function, a major addition in SMT compared to SAT \cite{tacas08-demoura-bjorner-z3-smt-solver, lin_scalable_2023}.
Concretely, the only set of constraints follows from \autoref{eq:def}: for every pair of distinct 1's at $(i,j)$ and $(i',j')$,
\begin{equation}
\biggl\{
\begin{array}{ll}
    f_{i,j} \neq f_{i',j'} & \text{if } M_{i,j'}=0, \\
\left(f_{i,j} = f_{i',j'}\right) \Rightarrow \left(f_{i,j} = f_{i,j'}\right) & \text{if } M_{i,j'}=1.
\end{array}
\end{equation}
Another SMT feature we leverage are bit-vectors.
In fact, both the domain and range of $f$ are bit-vectors: $f_{i,j}$ above means $f(e(i,j))$ where $e$ is an index function of the 1's in $M$, and the value of $f$ is the index of a rectangle.
To narrow down the solution space as in line 8 of \autoref{alg}, we just add new constraints $f_{i,j}\neq b$ for every $M_{i,j}=1$ to the SMT formula.

\subsection{Heuristics \label{ssec:heuristics}}

A trivial upper bound of $r_\mathbb{B}(M)$ is the width or height of $M$, whichever smaller, after removing empty and duplicated rows and columns.
This corresponds to partitioning the matrix into single rows or columns and consolidating duplicated ones.

\begin{figure*}[t]
\centerline{\includegraphics[scale=1.3]{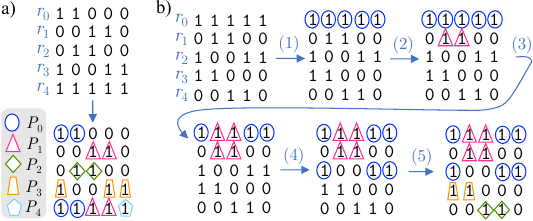}}
\caption{Two trials of running the row packing heuristic.
Rectangles found are represented by different markers. a) needs 5 rectangles but b) needs 4.}
\label{fig:eg}
\end{figure*}

The normal set basis viewpoint inspires our second heuristic.
We process matrix $M$ row by row, with the goal of forming a \textit{basis} -- each basis vector corresponds to one rectangle, as outlined in \autoref{alg:packing}.
For each row $r_i$, as in lines 4-7, if an existing basis vector $v_j$ is found within this row, we append $i$ to the rectangle $P_j$ associated with $v_j$.
Subsequently, we remove the 1's in $v_j$ from $r_i$ and continue this process.
The outcome is the decomposition of $r_i$ into a disjoint union of existing basis vectors, potentially leaving a \textit{residue} of 1's.
An example is displayed in \autoref{fig:eg}a where the first 4 rows cannot be decomposed, so the residues are just the rows themselves, and they are added to the basis, i.e., $v_i=r_i,\ i\in\{0,1,2,3\}$.
When it comes to $r_4$, we note it contains $v_0$ (circles) and $v_1$ (triangles), so the residue is $(0,0,0,0,1)$ denoted by the pentagon.
Based on this decomposition, the rectangles $P_0$ and $P_1$, corresponding to $v_0$ and $v_1$, vertically grow to include row 4.

Since we adhere to the order of basis vectors, the decomposition can be suboptimal.
For instance, we overlook the possibility of $r_4=v_2+v_3$, and the residue could have been 0.
To mitigate this, we run the heuristic multiple times, shuffling the rows in each trial, e.g., another trial with a different row ordering is exhibited in \autoref{fig:eg}b.
This is a compromise to the complexity of the problem.
Formally, we are trying to find a \textit{packing} or \textit{exact cover} of $r_i$ by the basis vectors, and it is an NP-complete problem~\cite{karp_reducibility_1972} to decide whether one exists.

\begin{algorithm}[t]
\caption{Row-Packing EBMF}\label{alg:packing}
\KwData{$M\in \mathbb{B}^{m\times n}$}
\KwResult{$P$, an EBMF of $M$ consisting of rectangles}
$M' \gets$ shuffle\_rows($M$)\;
\textit{basis} $\gets [\ ]$; \ \ $P \gets[\ ]$\;
\For{$r_i \in M'\ \ i=0,1,...,m-1$}{
  \For{$v_j\in$ basis}{
    \If{ \{1's in $v_j$\} $\subseteq$ \{1's in $r_i$\} }{
      $P_j\gets$ vertical\_grow($P_j$, $i$)\;
      $r_i \gets r_i-v_j$\;
    }
  }
  \If{$r_i \neq \vec 0$}{
  $c\gets$ one\_hot\_column\_vec($i$)\;
  \For{$v_k\in$ basis}{
    \If{ \{1's in $r_i$\} $\subseteq$ \{1's in $v_k$\} }{
      $P_k\gets$ horizontal\_shrink($P_k$, $r_i$)\;
      $v_k \gets v_k-r_i$\;
      $c_k \gets 1$\;
    }
  }
    \textit{basis}.append(\textit{$r_i$})\;
    $P$.append($c \times r_i$)\;
  }
}
$P\gets$ undo\_shuffle($P$, $M$, $M'$)\;
\end{algorithm}

In the presence of a residue, we perform an update to the basis in lines 9-16.
The intuition behind this is that smaller basis vectors enhance the likelihood of a successful row packing.
If an existing basis vector $v_k$ contains the residue, we remove the residue from the corresponding rectangle $P_k$ and update $v_k$.
In step 2 of \autoref{fig:eg}b, $r_1$ itself (triangles) is the residue. 
We find existing basis vector $v_0=r_0$ (circles) containing $r_1$, so $v_0$ is updated to $r_0-r_1$, and then $r_1$ is added to the basis as $v_1$.
Because of the updates, some existing rectangles shrink to remove the columns of 1's in the new basis vector, which we record with a column vector $c$.
In step 3 of the example, $c$ notes that $v_0$ gets updated and $v_1$ gets added, so the new rectangle $P_1=c\cdot v_1$ spans rows 0 and 1, and columns 0 and 1.
And the existing rectangle $P_0$ shrinks by removing columns 0 and 1, which leads to the successful packing of $r_2$ in step 4.

Finally, in line 17, we reverse the initial shuffling to derive the correct EBMF.
It is worth noting that the algorithm introduces at most one rectangle for each non-repeating row, ensuring that the result is no worse than the trivial heuristic.
The overall time complexity is $O(n^3k)$, where $k$ represents the number of trials, and $n$ denotes the larger of matrix width and height.
This complexity is due to nested loops in lines (3,10) and (3,4), with the innermost loop involving vector operations.
Additionally, note that we run the heuristic on both the original matrix and its transpose, retaining the better result.

In scenarios with limited time budgets, two further compromises can be considered.
The first one is removing the basis update in lines 10-14.
The second one is arranging rows with a smaller number of 1's at the beginning instead of random shuffling, and invoking fewer runs.
Based on our experience, both of these tend to result in more suboptimal `local minima' compared to the current setting, so we have not adopted them.

\section{Evaluation \label{sec:evaluation}}

We implement the above approach which is open-source under the MIT License\footnote{\url{https://github.com/UCLA-VAST/EBMF}}.
The software relies on numpy 1.26.3 and z3-solver 4.12.1.0~\cite{tacas08-demoura-bjorner-z3-smt-solver}.
The evaluation is conducted on a server with an AMD EPYC 7V13 CPU and 512 GB RAM.

\subsection{Benchmark Construction}

We provide benchmarks in two sizes: 1) limiting the number of rows by 10 so that we can reliably prove the optimality of the solutions using SMT, and 2) 100$\times$100, which is considered to be the current limit of atom array technology~\cite{bluvstein_logical_2023}.

The first benchmark set consists of random matrices.
We generate 10 matrices with varying occupancies of 1's (10\%, 20\%, ..., 90\%) for sizes 10$\times$10, 10$\times$20, and 10$\times$30.
For the 100$\times$100 size, we choose occupancies of 1\%, 2\%, 5\%, 10\%, and 20\%, because higher occupancies almost always result in full rank, which is trivial for our evaluation.

The second benchmark set is comprised of matrices with known optimal solutions.
According to \autoref{eq:lowerbound}, if a matrix has a $k$-rectangle partition and the real rank is also $k$, the partition is optimal.
We create pairs of disjoint rows $r_i$ and linearly independent columns $c_i$, leading to matrices $M=\sum_{i=1}^k c_i\cdot r_i$.
We enforce disjointedness among the rows to ensure that the outcome matrices only contain 0's and 1's, and the rectangles cannot merge.
For each rank $k=1,2,...,10$, we generate 10 benchmarks of size 10$\times$10 with known optimal solutions.

The third benchmark set is designed to create a gap between the real rank and the binary rank.
We begin by sampling a random row $r$ and then randomly decompose it into disjoint row pairs $r=r'+r''$.
The parameter for this family of benchmarks is the number of row pairs, $k$, which is limited to $\lfloor m/2\rfloor$.
The real rank of these $2k$ rows should be $k+1$ because any pair can recover the original row, $r=r_0+r_1$.
Each pair then should provide an independent basis vector, e.g., $r_{2i}$ for $i=0,...,k-1$. 
Note that decompositions like $r_3=r_0+r_1-r_2$ require the use of negative numbers, which are not allowed in an EBMF.
Consequently, the binary rank of the matrix should be larger than $k+1$.
The remaining $m-2k$ rows are completed with random rows having a 50\% occupancy, resulting in a total real rank equal to or slightly lower than $m-k+1$.
We generate 100 benchmarks of size 10$\times$10 with 2, 3, 4, and 5 row pairs.

\subsection{Results}

The SMT solver allows us to compute optimal solutions.
The percentage of cases achieving optimal solutions with the heuristics is presented in \autoref{tab}.
The `rank' column indicates the percentage of cases where the binary rank equals the real rank.
Although the 100$\times$100 benchmarks are too large for SMT to find solutions, the heuristics find solutions with the number of rectangles equal to the real rank.
Consequently, these solutions are known to be optimal, and the real and binary ranks are in agreement.
Several observations can be made.

\textit{Observation 1: the real and binary ranks are equal with high probability for random matrices.}
This can be attributed in part to the near-full real rank of random matrices.
We observe that almost all 10$\times$20 matrices with an occupancy of 20\% and higher, all 10$\times$30 matrices, and 100$\times$100 matrices with an occupancy of 5\% and higher are full rank, necessitating full binary rank, resulting in equality between the two.

\textit{Observation 2: the constructed benchmarks with known optimal are easy.}
Due to the mechanism of row packing, it can always find the optimal solutions for these benchmarks.
Surprisingly, the trivial heuristic also manages to find the optimal solutions on all cases, because even though the row space cannot be reduced by construction, e.g.,
\begin{equation*}
    \begin{bmatrix}
        1 \\ 1 \\ 0
    \end{bmatrix} \begin{bmatrix}
        1 & 1 & 0
    \end{bmatrix} +
    \begin{bmatrix}
        0 \\ 1 \\ 1
    \end{bmatrix}
    \begin{bmatrix}
        0 & 0 & 1
    \end{bmatrix}=
    \begin{bmatrix}
       1 & 1 & 0 \\
       1 & 1 & 1 \\
       0 & 0 & 1
    \end{bmatrix},
\end{equation*}
the columns may be reduced by recognizing duplication.

\textit{Observation 3: row packing is an effective heuristic.}
On benchmarks with gaps and the large random benchmarks, there is a big gap between the trivial heuristic and even one trial of row packing, indicating row packing is highly non-trivial.
As expected, the performance of row packing improves with more trials.
On most of the benchmarks, it saturates at 100 trials and finds optimal solutions on a remarkable percentage of cases.

\begin{table}[t]
\caption{Percentage of cases finding an optimal solution}
\begin{center}
\begin{tabular}{|c|c|c|c|c|c|c|}
\hline
 & & & \multicolumn{4}{c|}{\textbf{row packing, number of trials}} \\
\cline{4-7} 
\textbf{benchmark} & \textbf{rank$^\dagger$} & \textbf{trivial} &\textbf{1} & \textbf{10} & \textbf{100} & \textbf{1000} \\
\hline
10$\times$10, rand & 98\% & 80\% & 91\% & 99\% & 100\% & 100\% \\
10$\times$20, rand & 100\% & 100\% & 100\% & 100\% & 100\% & 100\% \\
10$\times$30, rand & 100\% & 100\% & 100\% & 100\% & 100\% & 100\%  \\
100$\times$100, rand & 100\%$^\ddagger$ & 62\% & 92\% & 96\% & 98\% & 100\% \\
10$\times$10, opt & 100\% & 100\% & 100\% & 100\% & 100\% & 100\%  \\
10$\times$10, gap, 2 & 74\% & 29\% & 88\% & 100\% & 100\% & 100\% \\
10$\times$10, gap, 3 & 63\% & 16\% & 91\% & 100\% & 100\% & 100\% \\
10$\times$10, gap, 4 & 47\% & 40\% & 94\% & 98\% & 99\% & 99\% \\
10$\times$10, gap, 5 & 42\% & 84\% & 90\% & 94\% & 96\% & 96\% \\
\hline
\end{tabular}
\label{tab}
\end{center}
\footnotesize{
$^\dagger$Percentage of cases where real and binary ranks are the same. \\
$^\ddagger$Since the heuristics managed to find optimal solutions, the real and binary ranks are the same for this set. The SMT for these cases are too large to solve.}
\end{table}

\begin{figure}[t]
\centerline{\includegraphics[scale=1]{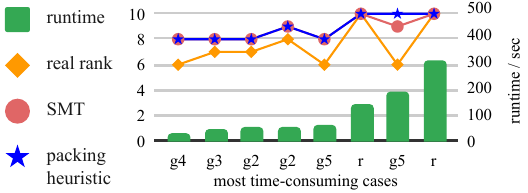}}
\caption{The most time-consuming cases. `r' means it is a random benchmark, `g2' means it comes from benchmarks with gap using 2 row pairs, etc.}
\label{fig:eva}
\end{figure}

\textit{Observation 4: edge cases for row packing needs more general search.}
We look into the cases where row packing fails to find the optimal solution.
Going through \autoref{alg:packing}, we find these cases necessitates introducing more than one basis at some rows, whereas the row packing heuristic at most introduces one new basis per row in order for efficiency.

\textit{Observation 5: the most time consuming cases are proving UNSAT.}
We collect the most time-consuming cases in \autoref{fig:eva}.
In the majority of these cases, the SMT solver can only find solutions with the same number of rectangles as row packing.
Then, the solver goes on decreasing the bound by 1 and proves the formula to be UNSAT.
This is the most time consuming task.
Note that in \autoref{alg}, when we terminate at any time, we can return $P$, the best solution found so far.

\section{Fault-Tolerant Quantum Computing \label{sec:ftqc}}
\begin{figure}[t]
\centerline{\includegraphics[scale=1]{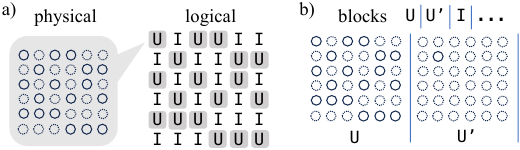}}
\caption{Rectangular addressing in fault-tolerant quantum computing.
a) An operation $U$ on 2D patterns of logical qubits can be realized by the tensor product of partitions on the logical and physical levels.
b) For logical blocks in 1D layout and with different operations, addressing by row is usually enough.}
\label{fig:ftqc}
\end{figure}

Fault-tolerant quantum computing performs on top of quantum error correction codes that encode each logical qubit using quantum states distributed across multiple physical qubits.
A promising approach is exemplified by the surface code~\cite{defect-surface-code}, where a logical qubit manifests as a patch of physical qubits, as depicted in \autoref{fig:ftqc}a.
For simplicity, only the data qubits are illustrated, and check qubits are not shown.
A single-logical-qubit operation, designated as $U$, corresponds to a 2D pattern ($M$) of physical gates, as highlighted in the callout.
On the logical level, the quantum circuit may necessitate another 2D pattern ($\hat M$) of logical operations.
Consequently, the overall physical operation is expressed as the tensor product $\hat M \otimes M$.
This two-level structure allows for the independent computation of the rectangular partition of $\hat M$ and $M$.
Subsequently, taking the tensor product of the partitions produces the solution.

However, is this solution optimal?
The real rank is multiplicative under a tensor product, as elementary row operations can be employed to make both $M$ and $\hat M$ upper triangular, resulting in an upper-triangular tensor product.
In contrast, whether the binary rank is multiplicative under a tensor product remains an open question.
Our aforementioned solution (tensor product of partitions) provides an upper bound: $r_\mathbb{B}(\hat M \otimes M)\le r_\mathbb{B}(\hat M)\cdot r_\mathbb{B}(M)$.
For lower bounds, Watson~\cite{watson_nonnegative_2016} notes that
\begin{equation}
\max\left(r_\mathbb{B}(\hat M) \cdot \phi(M),\ r_\mathbb{B}(M) \cdot \phi(\hat M) \right) \le r_\mathbb{B}(\hat M \otimes M)
\end{equation}
where $\phi$ denotes the maximum fooling set size.
However, as per \autoref{eq:fooling}, $\phi$ is not always equal to $r_\mathbb{B}$.
In practice, the majority of $M$ is simple, such as applying $X$, $Z$, or $H$ to all the physical qubits in one patch.
In this case, all the elements of $M$ are 1, and indeed we have $\phi(M)=r_\mathbb{B}(M)=1$, so the rectangular partition of $\hat M$ leads to an optimal solution.

Another family of quantum error correction codes gaining popularity is quantum low-density parity-check codes, which can take advantage of the mobility of atom arrays~\cite{xu_constant-overhead_2023}.
In this code, logical qubits are more globalized, with multiple logical qubits stored in one \textit{logical block} instead of one qubit per block.
These blocks are usually arranged in a 1D fashion, as shown in \autoref{fig:ftqc}b, because they only serve as memory, and logical qubits need to be read out to a computing zone.
Considering logical operations that can be realized with single-qubit gates in this setting, the pattern on each block can be quite different, depending on the offset of logical qubits inside the blocks.
We conjecture that addressing qubits row by row is usually sufficient in this case, as in our evaluation, we find that given the same occupancy, the 10$\times$20 and 10$\times$30 random matrices are much easier to be full rank than the 10$\times$10 matrices.

\section{Conclusion and Discussion\label{sec:conclusion}}
In this paper, we consider the depth-optimal rectangular addressing for 2D qubit arrays, which turns out to be equivalent to the \textit{exact binary matrix factorization} problem that has applications in various fields.
We introduce an SMT-based solver for it along with an effective heuristic, \textit{row packing}, which scales to the current limits of atom array technology.
Several future directions can be explored.
The packing procedure in our implementation might benefit from ideas in existing works such as Knuth's Algorithm X for exact cover~\cite{knuth_dancing_2000} instead of purely relying on shuffling. 
Another avenue is the introduction of vacancies in the atom arrays.
Since there are no qubits, it is irrelevant how many times we address these sites.
They can be represented as \textit{don't cares} in a matrix, which may be leveraged to reduce rectangles.
This task is binary matrix \textit{completion}~\cite{beckerleg_divide-and-conquer_2020, yadava_boolean_nodate} instead of factorization.
Additionally, the SMT tool could aid in investigating the behavior of binary rank under a tensor product.
Furthermore, optimal rectangular partitions generated by this tool can provide insight for designing cryogenic control architecture of qubit arrays for various applications.

\section*{Acknowledgement}
This work is funded by NSF grant 442511-CJ-22291.
The authors would like to thank Y. Song for discussions on superconducting circuits, D. Bluvstein and H. Zhou for conversations on neutral atom arrays, and a post on TCS Stack Exchange about binary rank by D. Issac, R. Kothari, and S. Nikolov.

\bibliographystyle{IEEEtranS}
\bibliography{refs_ori}

\begin{thebibliography}{10}
\providecommand{\url}[1]{#1}
\csname url@samestyle\endcsname
\providecommand{\newblock}{\relax}
\providecommand{\bibinfo}[2]{#2}
\providecommand{\BIBentrySTDinterwordspacing}{\spaceskip=0pt\relax}
\providecommand{\BIBentryALTinterwordstretchfactor}{4}
\providecommand{\BIBentryALTinterwordspacing}{\spaceskip=\fontdimen2\font plus
\BIBentryALTinterwordstretchfactor\fontdimen3\font minus \fontdimen4\font\relax}
\providecommand{\BIBforeignlanguage}[2]{{%
\expandafter\ifx\csname l@#1\endcsname\relax
\typeout{** WARNING: IEEEtranS.bst: No hyphenation pattern has been}%
\typeout{** loaded for the language `#1'. Using the pattern for}%
\typeout{** the default language instead.}%
\else
\language=\csname l@#1\endcsname
\fi
#2}}
\providecommand{\BIBdecl}{\relax}
\BIBdecl

\bibitem{bluvstein_logical_2023}
D.~Bluvstein, S.~J. Evered, A.~A. Geim, S.~H. Li, H.~Zhou, T.~Manovitz, S.~Ebadi, M.~Cain, M.~Kalinowski, D.~Hangleiter, J.~Pablo, B.~Ataides, N.~Maskara, I.~Cong, X.~Gao, P.~S. Rodriguez, T.~Karolyshyn, G.~Semeghini, M.~Gullans, M.~Greiner, V.~Vuletić, and M.~D. Lukin, ``Logical quantum processor based on reconfigurable atom arrays,'' \emph{Nature}, vol. 626, pp. 58 -- 65, 2023, \href{https://doi.org/10.1038/s41586-022-04592-6}{DOI: \detokenize{10.1038/s41586-022-04592-6}}.

\bibitem{nature22-graham-atom-array}
T.~M. Graham, Y.~Song, J.~Scott, C.~Poole, L.~Phuttitarn, K.~Jooya, P.~Eichler, X.~Jiang, A.~Marra, B.~Grinkemeyer, M.~Kwon, M.~Ebert, J.~Cherek, M.~T. Lichtman, M.~Gillette, J.~Gilbert, D.~Bowman, T.~Ballance, C.~Campbell, E.~D. Dahl, O.~Crawford, N.~S. Blunt, B.~Rogers, T.~Noel, and M.~Saffman, ``\BIBforeignlanguage{en}{Multi-qubit entanglement and algorithms on a neutral-atom quantum computer},'' \emph{\BIBforeignlanguage{en}{Nature}}, vol. 604, no. 7906, pp. 457--462, Apr. 2022, \href{https://doi.org/10.1038/s41586-022-04603-6}{DOI: \detokenize{10.1038/s41586-022-04603-6}}.

\bibitem{ebadi_quantum_2021}
S.~Ebadi, T.~T. Wang, H.~Levine, A.~Keesling, G.~Semeghini, A.~Omran, D.~Bluvstein, R.~Samajdar, H.~Pichler, W.~W. Ho, S.~Choi, S.~Sachdev, M.~Greiner, V.~Vuletić, and M.~D. Lukin, ``\BIBforeignlanguage{en}{Quantum phases of matter on a 256-atom programmable quantum simulator},'' \emph{\BIBforeignlanguage{en}{Nature}}, vol. 595, no. 7866, pp. 227--232, Jul. 2021, \href{https://doi.org/10.1038/s41586-021-03582-4}{DOI: \detokenize{10.1038/s41586-021-03582-4}}.

\bibitem{bluvstein_quantum_2022}
D.~Bluvstein, H.~Levine, G.~Semeghini, T.~T. Wang, S.~Ebadi, M.~Kalinowski, A.~Keesling, N.~Maskara, H.~Pichler, M.~Greiner, V.~Vuletić, and M.~D. Lukin, ``\BIBforeignlanguage{en}{A quantum processor based on coherent transport of entangled atom arrays},'' \emph{\BIBforeignlanguage{en}{Nature}}, vol. 604, no. 7906, pp. 451--456, Apr. 2022, \href{https://doi.org/10.1038/s41586-022-04592-6}{DOI: \detokenize{10.1038/s41586-022-04592-6}}.

\bibitem{tan_qubit_2022}
B.~Tan, D.~Bluvstein, M.~D. Lukin, and J.~Cong, ``\BIBforeignlanguage{en}{Qubit mapping for reconfigurable atom arrays},'' in \emph{\BIBforeignlanguage{en}{Proceedings of the 41st {IEEE}/{ACM} {International} {Conference} on {Computer}-{Aided} {Design}}}, Oct. 2022, \href{https://doi.org/10.1145/3508352.3549331}{DOI: \detokenize{10.1145/3508352.3549331}}.

\bibitem{Tan2024compilingquantum}
D.~B. Tan, D.~Bluvstein, M.~D. Lukin, and J.~Cong, ``Compiling quantum circuits for dynamically field-programmable neutral atoms array processors,'' \emph{{Quantum}}, vol.~8, p. 1281, Mar. 2024, \href{https://doi.org/10.22331/q-2024-03-14-1281}{DOI: \detokenize{10.22331/q-2024-03-14-1281}}.

\bibitem{yao_complexity_1979}
A.~C.-C. Yao, ``Some complexity questions related to distributive computing,'' in \emph{Proceedings of the eleventh annual {ACM} symposium on {Theory} of computing}, New York, NY, USA, Apr. 1979, pp. 209--213, \href{https://doi.org/10.1145/800135.804414}{DOI: \detokenize{10.1145/800135.804414}}.

\bibitem{kushilevitz_communication_1997}
E.~Kushilevitz and N.~Nisan, \emph{\BIBforeignlanguage{en}{Communication Complexity}}.\hskip 1em plus 0.5em minus 0.4em\relax Cambridge University Press, 1997, ch. 1.1, \href{https://doi.org/10.1017/CBO9780511574948}{DOI: \detokenize{10.1017/CBO9780511574948}}.

\bibitem{jiang_minimal_1993}
T.~Jiang and B.~Ravikumar, ``Minimal {NFA} problems are hard,'' \emph{SIAM Journal on Computing}, vol.~22, no.~6, pp. 1117--1141, Dec. 1993, \href{https://doi.org/10.1137/0222067}{DOI: \detokenize{10.1137/0222067}}.

\bibitem{bein_clustering_2008}
D.~Bein, L.~Morales, W.~Bein, J.~C.~O.~Shields, Z.~Meng, and I.~H. Sudborough, ``\BIBforeignlanguage{English}{Clustering and the biclique partition problem},'' in \emph{\BIBforeignlanguage{English}{Proceedings of the 41st Annual Hawaii International Conference on System Sciences}}, Jul. 2008, \href{https://doi.org/10.1109/HICSS.2008.504}{DOI: \detokenize{10.1109/HICSS.2008.504}}.

\bibitem{chalermsook_nearly_2014}
P.~Chalermsook, S.~Heydrich, E.~Holm, and A.~Karrenbauer, ``\BIBforeignlanguage{en}{Nearly tight approximability results for minimum biclique cover and partition},'' in \emph{\BIBforeignlanguage{en}{Algorithms - European Symposium on Algorithms 2014}}, 2014, pp. 235--246, \href{https://doi.org/10.1007/978-3-662-44777-2\_20}{DOI: \detokenize{10.1007/978-3-662-44777-2\_20}}.

\bibitem{chandran_parameterized_2017}
S.~Chandran, D.~Issac, and A.~Karrenbauer, ``On the parameterized complexity of biclique cover and partition,'' in \emph{11th International Symposium on Parameterized and Exact Computation}, 2016, \href{https://doi.org/10.4230/LIPICS.IPEC.2016.11}{DOI: \detokenize{10.4230/LIPICS.IPEC.2016.11}}.

\bibitem{amilhastre_complexity_1998}
J.~Amilhastre, M.~Vilarem, and P.~Janssen, ``\BIBforeignlanguage{en}{Complexity of minimum biclique cover and minimum biclique decomposition for bipartite domino-free graphs},'' \emph{\BIBforeignlanguage{en}{Discrete Applied Mathematics}}, vol.~86, no. 2-3, pp. 125--144, Sep. 1998, \href{https://doi.org/10.1016/S0166-218X(98)00039-0}{DOI: \detokenize{10.1016/S0166-218X(98)00039-0}}.

\bibitem{watson_nonnegative_2016}
T.~Watson, ``\BIBforeignlanguage{en}{Nonnegative rank vs. binary rank},'' \emph{\BIBforeignlanguage{en}{Chicago Journal of Theoretical Computer Science}}, vol.~22, no.~1, 2016, \href{https://doi.org/10.4086/cjtcs.2016.002}{DOI: \detokenize{10.4086/cjtcs.2016.002}}.

\bibitem{zhang_binary_2007}
Z.~Zhang, T.~Li, C.~Ding, and X.~Zhang, ``\BIBforeignlanguage{en}{Binary matrix factorization with applications},'' in \emph{\BIBforeignlanguage{en}{Seventh {IEEE} {International} {Conference} on {Data} {Mining}}}, Oct. 2007, pp. 391--400, \href{https://doi.org/10.1109/ICDM.2007.99}{DOI: \detokenize{10.1109/ICDM.2007.99}}.

\bibitem{Zitnik2012}
M.~Zitnik and B.~Zupan, ``{NIMFA}: A {Python} library for nonnegative matrix factorization,'' \emph{Journal of Machine Learning Research}, vol.~13, pp. 849--853, 2012, \href{https://doi.org/10.48550/arXiv.1808.01743}{DOI: \detokenize{10.48550/arXiv.1808.01743}}.

\bibitem{tacas08-demoura-bjorner-z3-smt-solver}
L.~{de Moura} and N.~Bj{\o}rner, ``\BIBforeignlanguage{en}{{Z3}: An efficient {SMT} solver},'' in \emph{\BIBforeignlanguage{en}{Tools and Algorithms for the Construction and Analysis of Systems}}, C.~R. Ramakrishnan and J.~Rehof, Eds., 2008, pp. 337--340, \href{https://doi.org/10.1007/978-3-540-78800-3\_24}{DOI: \detokenize{10.1007/978-3-540-78800-3\_24}}.

\bibitem{lin_scalable_2023}
W.-H. Lin, J.~Kimko, B.~Tan, N.~Bjørner, and J.~Cong, ``\BIBforeignlanguage{en}{Scalable optimal layout synthesis for {NISQ} quantum processors},'' in \emph{\BIBforeignlanguage{en}{2023 60th {ACM}/{IEEE} {Design} {Automation} {Conference}}}, Jul. 2023, \href{https://doi.org/10.1109/DAC56929.2023.10247760}{DOI: \detokenize{10.1109/DAC56929.2023.10247760}}.

\bibitem{karp_reducibility_1972}
R.~M. Karp, ``\BIBforeignlanguage{en}{Reducibility among combinatorial problems},'' in \emph{\BIBforeignlanguage{en}{Complexity of Computer Computations}}, ser. The {IBM} Research Symposia Series, R.~E. Miller, J.~W. Thatcher, and J.~D. Bohlinger, Eds.\hskip 1em plus 0.5em minus 0.4em\relax Boston, MA: Springer, 1972, pp. 85--103, \href{https://doi.org/10.1007/978-1-4684-2001-2\_9}{DOI: \detokenize{10.1007/978-1-4684-2001-2\_9}}.

\bibitem{defect-surface-code}
A.~G. Fowler, M.~Mariantoni, J.~M. Martinis, and A.~N. Cleland, ``Surface codes: Towards practical large-scale quantum computation,'' \emph{Phys. Rev. A}, vol.~86, p. 032324, Sep 2012, \href{https://doi.org/10.1103/PhysRevA.86.032324}{DOI: \detokenize{10.1103/PhysRevA.86.032324}}.

\bibitem{xu_constant-overhead_2023}
Q.~Xu, J.~P.~B. Ataides, C.~A. Pattison, N.~Raveendran, D.~Bluvstein, J.~Wurtz, B.~Vasic, M.~D. Lukin, L.~Jiang, and H.~Zhou, ``Constant-overhead fault-tolerant quantum computation with reconfigurable atom arrays'', Aug. 2023, \href{https://arxiv.org/abs/2308.08648}{arXiv: quant-ph/2308.08648}.

\bibitem{knuth_dancing_2000}
D.~E. Knuth, ``\BIBforeignlanguage{en}{Dancing links},'' \emph{\BIBforeignlanguage{en}{Millenial Perspectives in Computer Science}}, pp. 187--214, Nov. 2000, \href{https://doi.org/10.48550/arXiv.cs/0011047}{DOI: \detokenize{10.48550/arXiv.cs/0011047}}.

\bibitem{beckerleg_divide-and-conquer_2020}
M.~Beckerleg and A.~Thompson, ``\BIBforeignlanguage{en}{A divide-and-conquer algorithm for binary matrix completion},'' \emph{\BIBforeignlanguage{en}{Linear Algebra and its Applications}}, vol. 601, pp. 113--133, Sep. 2020, \href{https://doi.org/10.1016/j.laa.2020.04.017}{DOI: \detokenize{10.1016/j.laa.2020.04.017}}.

\bibitem{yadava_boolean_nodate}
\BIBentryALTinterwordspacing
P.~Yadava, ``Boolean matrix factorization with missing values,'' Master's thesis, Universit\"at des Saarlandes, 2012. [Online]. Available: \url{https://hdl.handle.net/11858/00-001M-0000-0014-627A-3}
\BIBentrySTDinterwordspacing

\end{thebibliography}

\end{document}